# Chaotic Dynamics and Optical Power Saturation in Parity-Time (*PT*) Symmetric Double Ring Resonator

Jyoti Prasad Deka and Amarendra K. Sarma*

**Abstract** In this work, we report emergence of saturation and chaotic spiking of optical power in a double ring resonator with balanced loss and gain, obeying the so-called parity-time symmetry. We have modeled the system using a discrete-time iterative equation known as the Ikeda Map. In the linear regime, evolution of optical power in the system shows power saturation behavior below the *PT* threshold and exponential blow-up above the *PT* threshold. We found that in the unbroken *PT* regime, optical power saturation occurs owing to the existence of stable stationary states, which lies on the surface of 4-dimensional hypersphere. Inclusion of Kerr nonlinearity into our model leads to the emergence of a stable, chaotic and divergent region in the parameter basin for period-1 cycle. A closer inspection into the system shows us that the largest Lyapunov exponent blows up in the divergent region. It is found that the existence of high non-negative largest Lyapunov exponent causes chaotic spiking of optical power in the resonators.

Department of Physics, Indian Institute of Technology Guwahati, Guwahati - 781039, Assam, India
*Corresponding author: aksarma@iitg.ernet.in

## 1 Introduction

Since the pioneering work of Bender and Boettcher [1], which showed that a wide class of non-Hermitian Hamiltonians can exhibit entirely real spectra as long as they respect the conditions of parity and time (PT) symmetry, research in non-Hermitian systems has virtually exploded in recent years. Non-Hermitian Hamiltonians undergo phase transition in some specific parametric regime and spontaneous breakdown of their eigenspectra occurs at the exceptional point (EP), beyond which the eigenspectra cease to be real. The exceptional point is also known as the *PT* threshold.

In recent times, *PT* symmetry has attracted considerable attention in optics and photonics. Christodoulide's group [2] proposed the idea that evanescently coupled dielectric optical waveguide with balanced gain and loss could provide a possible way for the experimental realization of *PT* symmetry in optical systems. The mathematical isomorphism between the time independent Schrodinger equation in quantum mechanics, and the paraxial equation of diffraction in optics made such a proposition possible. A. Guo *et al.* [3] and Ruter *et al.* [4] soon reported the first experimental observation of *PT* symmetry in a coupled optical waveguide system. Since then, innumerable works have been reported in theoretical as well as experimental settings. To cite, some of them include onset of chaos in optomechanical systems [5], Peregrine soliton dynamics [6], whispering gallery modes [7], optical oligomers [8-10] and so on. Lately, another area where PT-symmetry related ideas are explored is nonlinear dynamics [11]. In this regard, optical ring resonator systems are getting immense attention [12,13]. It is worthwhile to note that, K. Ikeda [14, 15] first studied the emergence of chaotic dynamics in such optical resonators. Ikeda's model studied the evolution dynamics of light going around a ring cavity with a nonlinear medium using a discrete-time iterative equation known as the Ikeda Map. Over the years, period doubling route to chaos [16], modulational instability [17], dynamical pulse shaping [18], temporal instabilities [19], etc. are some of the phenomena that have been studied theoretically as well as experimentally in such systems. Nonlinear dynamics of micro electromechanical mirror [20] and nano-mechanical membrane [21] have also been already studied in optical resonators. On the other hand, *PT* symmetry breaking induced chaotic dynamics [5] has been studied in cavity optomechanics. They found that chaos emerges in the system when the driving laser applied to the cavity mode lasts for a period. Moreover, in an earlier work, we studied the controllability of chaotic dynamics in a single fiber resonator [22]. We studied the system by modeling the evolution of optical power using the transfer matrix formalism [23]. We found out that the dissipation in the ring resonator and the input field both play major role in the emergence of chaotic dynamics in the system. In this article, we study a periodically driven double ring resonator system with balanced loss and gain. The ring resonators are coupled by a 50:50 passive directional coupler. One of the ring resonators is an amplifying resonator and the other is a lossy resonator, making the system parity-time symmetric. It may be noted that in our previous work [22], we studied a single lossy ring resonator coupled to an amplifying fiber via a passive 50:50 directional coupler. We observed controllable chaotic dynamics in the evolution of the resonator optical power. However, in spite of the system having balanced loss and gain, the system was not PT-symmetric.

The article is organized as follows. In Sect. 2 the discrete-time iterative equation using the transfer matrix formalism is worked out. Sect. 3 contains results and discussion. The *PT* threshold is evaluated in the linear regime and the dynamical behavior of the system in the unbroken *PT* regime and broken *PT* regime is analyzed. Then, the emergence of chaotic dynamics in the nonlinear regime is discussed. Finally, in Sect. 4 we conclude our work.

## 2 Theoretical Model

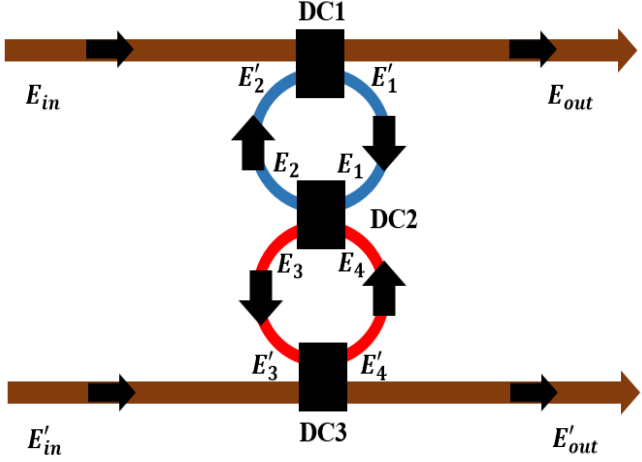

Fig. 1. Schematic diagram of the double ring resonator. The *black-color* block represents the 50:50 passive directional couplers (DC) and the *grey-color* region represents the optical fibers. The lossy resonator is colored *blue* while the amplifying resonator is colored *red*. $E_{in}$ and $E'_{in}$ are the input field amplitudes and $E_{out}$ and $E'_{out}$ are the output field amplitudes. $E'_1$ and $E'_4$ are the output field amplitudes of coupler region DC1 and DC3 respectively. They are transferred via the resonators to the coupler region DC2 as $E_1$ and $E_4$. Similarly, $E_2$ and $E_3$ are the output field amplitudes of coupler region DC2. They are transferred, via the resonators, to the coupler region DC1 and DC3 as $E'_2$ and $E'_3$.

The schematic of the model is shown in Fig. 1. It consists of three passive 50:50 directional couplers and two optical fiber-based ring resonators with equal amount of gain and loss. In addition, the two ring resonators exhibit Kerr nonlinearity of equal strength. The resonators are coupled via the directional couplers and driven periodically by the input field amplitudes $E_{in}$ and $E'_{in}$ as explained in Fig. 1.

Using the same formalism from our previous work [22], the evolution dynamics of the optical power in the resonator could be expressed as a discrete time iterative equation as follows.

$$E_{1,j+1} = e^{i\beta|E_{1,j}|^2}e^{-g/2}\left(iE_{in}/\sqrt{2} + e^{i\beta e^{-g}|E_{2,j}|^2}e^{-g/2}(E_{1,j} + iE_{4,j})/2\right) \quad (1)$$

Here, '$g$' is the gain/loss parameter and '$\beta$' is the strength of the Kerr nonlinear phase shift. Similarly, we can express the field amplitude $E_4$ in the form of a discrete-time iterative equation as given below:

$$E_{4,j+1} = e^{i\beta|E_{4,j}|^2}e^{g/2}\left(iE'_{in}/\sqrt{2} + e^{i\beta e^{g}|E_{3,j}|^2}e^{g/2}(iE_{1,j} + E_{4,j})/2\right) \quad (2)$$

In matrix notation, in the linear regime (i.e. $\beta = 0$) Eq. (1) and (2) could be expressed as follows:

$$\begin{pmatrix}E_{1,j+1}\\E_{4,j+1}\end{pmatrix} = T\begin{pmatrix}E_{1,j}\\E_{4,j}\end{pmatrix} + \frac{1}{\sqrt{2}}\begin{pmatrix}A_1\\A_2\end{pmatrix} \quad (3)$$

where $T = \begin{pmatrix}e^{-g}/2 & ie^{-g}/2\\ie^{g}/2 & e^{g}/2\end{pmatrix}$ is the transfer matrix. The driving terms are given by $A_1 = ie^{-g/2}E_{in}$ and $A_2 = ie^{g/2}E'_{in}$. To evaluate the *PT* threshold of the system, we solve for the eigenvalues of the matrix $T$ and find that the eigenvalues are given by: $\lambda = (\cosh(g) \pm \sqrt{(\cosh(g))^2 - 2})/2$. The *PT* threshold, above which the eigenvalues cease to be real, is found to be $g_{th} \approx 0.8814$.

## 3 Results and discussion

The evolution dynamics of optical power in the resonators could be investigated by solving Eq. (3). Fig. 2 depicts the optical powers, $P_1 = |E_1|^2$ and $P_4 = |E_4|^2$, in the resonators against the number of iterations. We find that the optical power saturation occurs when $g < g_{th}$ and $g = g_{th}$, i.e. below and at the PT-threshold, as evident from Fig. 2(a) and 2(b). On the other hand, as could be seen from Fig. 2(c), when $g > g_{th}$, i.e. above the PT-threshold, the optical power grows exponentially as the system evolves.

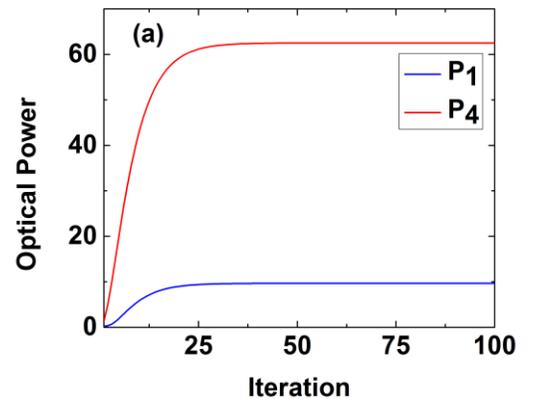

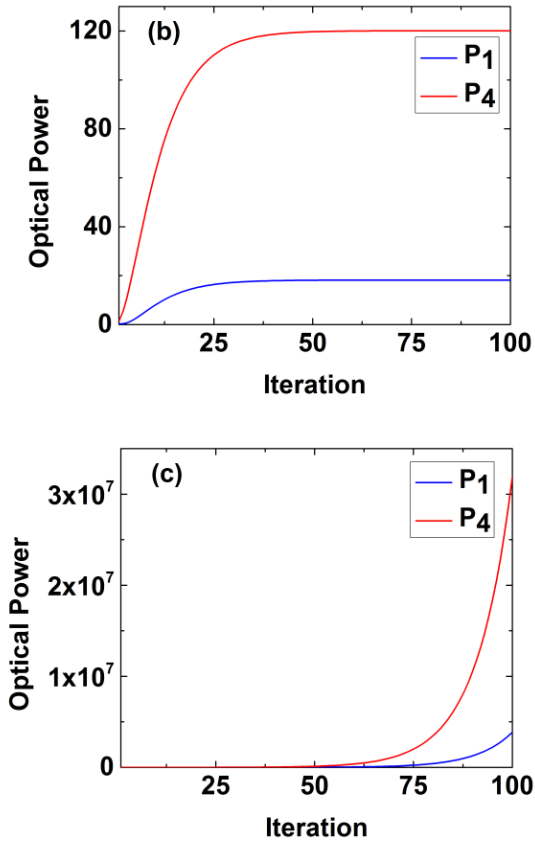

Fig. 2. Evolution of optical power in both the resonators. The parameters chosen are (a) $g = 0.85$ (Below the $PT$ Threshold) (b) (a) $g = 0.8814$ (At the $PT$ Threshold) (c) $g = 1$ (Above the $PT$ Threshold). The input field amplitudes are $E_{in} = 1$ and $E'_{in} = 1$ in all cases.

The onset of optical power saturation in the linear regime could be explained by constructing the hyper-sphere, on whose surface lies the stable stationary states of Eq. 3. The existence of stationary states could be ascertained by measuring the radius of the hyper-sphere. This will lead us to an indirect inference for the cause of optical power saturation in the system. By separating the field amplitudes $E_{1,j}$ and $E_{4,j}$ in Eq. 3, into its real and imaginary components, we obtain:

$$x_{1,j+1} = -A_1 y_{in} + B_1(x_{1,j} - y_{4,j}) \quad (4a)$$
$$y_{1,j+1} = A_1 x_{in} + B_1(x_{4,j} + y_{1,j}) \quad (4b)$$
$$x_{4,j+1} = -A_2 y'_{in} + B_2(x_{4,j} - y_{1,j}) \quad (4c)$$
$$y_{4,j+1} = A_2 x'_{in} + B_2(x_{1,j} + y_{4,j}). \quad (4d)$$

Here, $A_1 = e^{-g/2}/\sqrt{2}$, $B_1 = e^{-g}/2$, $A_2 = e^{g/2}/\sqrt{2}$, $B_2 = e^g/2$, $x_{i,j} = Re(E_{i,j})$ $y_{i,j} = Im(E_{i,j})$, $x_{in} = Re(E_{in})$, $y_{in} = Im(E_{in})$, $x'_{in} = Re(E'_{in})$ and $y'_{in} = Im(E'_{in})$. To evaluate the period-1 stationary states of Eq. 4, we set $x_{i,j+1} = x_{i,j}$ and $y_{i,j+1} = y_{i,j}$, where $i = 1,4$. Mathematical simplification leads us to the equation for a 4-dimensional hyper-sphere as follows:

$$(x_1 + A_1 C_1 y_{in})^2 + (y_1 - A_1 C_1 x_{in})^2 + (x_4 + A_2 C_2 y'_{in})^2 + (y_4 - A_2 C_2 x'_{in})^2 = A_1^2 D_1 P_{in} + A_2^2 D_2 P'_{in}$$

where $C_1 = (1 - B_1)/((1 - B_1)^2 - B_2^2)$, $C_2 = (1 - B_2)/((1 - B_2)^2 - B_1^2)$, $D_1 = B_2^2/((1 - B_1)^2 - B_2^2)^2$, $P_{in} = x_{in}^2 + y_{in}^2$, $D_2 = B_1^2/((1 - B_2)^2 - B_1^2)^2$ and $P'_{in} = {x'_{in}}^2 + {y'_{in}}^2$. The radius of the hypersphere is, $R = \sqrt{A_1^2 D_1 P_{in} + A_2^2 D_2 P'_{in}}$. The hypersphere could exist only if $R$ is of finite magnitude and a real number. The existence of the hypersphere would thereby prove why optical power saturation occurs below the $PT$ threshold.

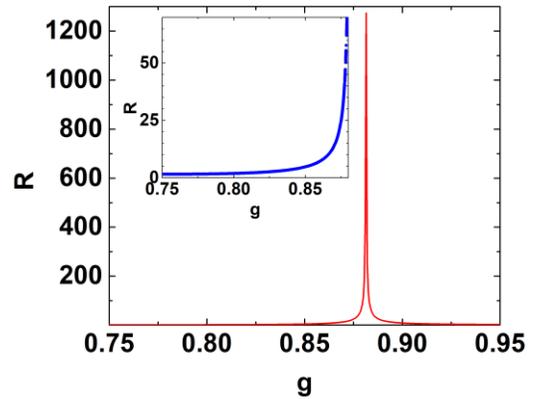

Fig. 3. Radius of the Hypersphere $R$ vs. Gain/Loss Coefficient $g$.

From Fig. 3, one could see that at $g = g_{th}$, $R$ increases to a very high magnitude, while beyond the $PT$ threshold, it decreases to zero sharply. However, in the unbroken $PT$ regime, it could be seen that the radius of the hypersphere is of finite magnitude. This means that the hypersphere exists only in the unbroken $PT$ regime and as such, it validates our reasoning as to why power saturation occurs. Next, we consider the case with Kerr nonlinearity, i.e. $\beta \neq 0$. Now, we find that the optical power evolution in the system is sensitive to the initial conditions. Here, the initial conditions are set by the excitation field amplitudes $E_{in}$ and $E'_{in}$. The parameter basin for period-1 cycle is exhibited in Fig. 4 for linear, weakly nonlinear and strongly nonlinear cases. In all the cases, we have taken the excitation field amplitude $E'_{in}$ as constant. The parameter basin consists of three regions namely, the convergent, the non-convergent and the divergent region, shown in blue, red and black color respectively in Fig. 4. In the convergent region, the optical power of the resonator will eventually converge to a stable stationary state. On the other hand, in the non-convergent region, it will oscillate in a period-$N$ cycle (where $N > 1$). However, in the divergent region, optical power in both resonators will blow up as the system evolves.

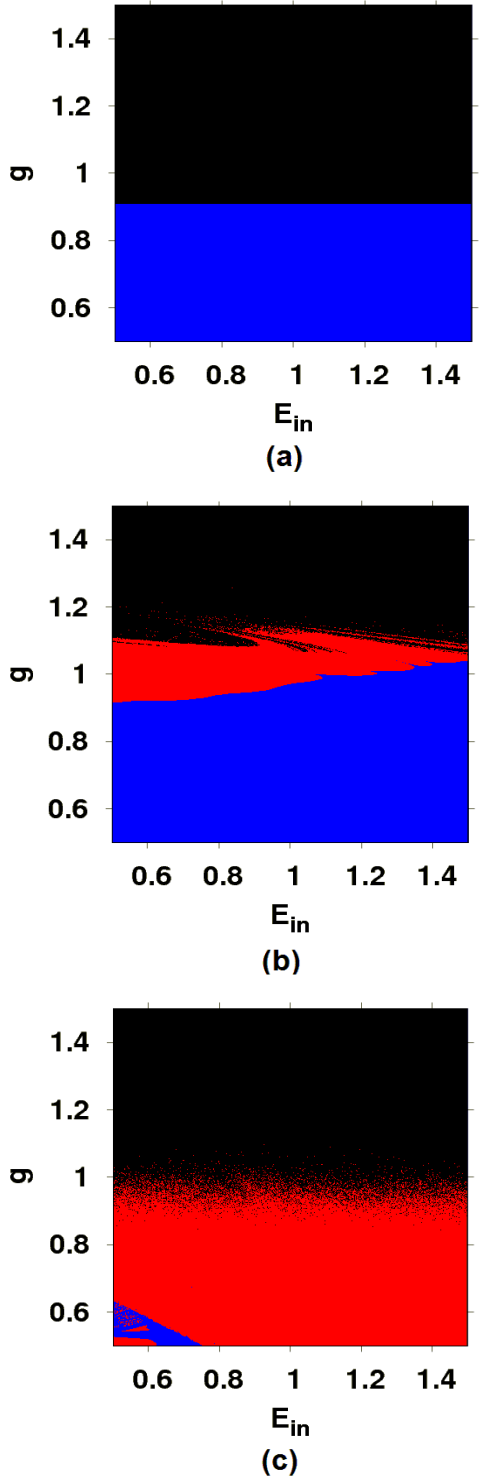

Fig. 4. Parameter Basin for Period-1 Cycle when $E'_{in} = 1$ and (a) $\beta = 0$ (Linear) (b) $\beta = 0.01$ (Weakly Nonlinear) (c) $\beta = 1$ (Strongly Nonlinear).

In the absence of nonlinearity, as could be seen from Fig. 4(a), we observe that the convergent and the divergent regions are easily discernible and there is absence of non-convergent region. This means that for $\beta = 0$, the system does not exhibit oscillatory behavior. However, this is not the case when we include nonlinearity into the system. In Fig. 4(b), the emergence of non-convergent region in the parameter basin could be observed. This implies the existence of period-$N$ cycle. In other words, it means that the optical power in both resonators will oscillate with every iteration. Moreover, when we consider the case of strong nonlinearity (Fig. 4(c)), the convergent region occupies a very small area in the parameter basin. Thus, one could infer that, as the strength of nonlinearity is increased, stable stationary states cease to exist in the system. Nevertheless, optical power in both resonators will blow-up as the gain/loss coefficient is enhanced beyond the PT threshold irrespective of the strength of nonlinearity.

Now, in order to have a clearer understanding, we study a specific case. The bifurcation diagram of optical power $P_1$ against the gain/loss parameter is depicted in Fig. 5(a), with the parameters: $E_{in} = 1$, $E'_{in} = 1$ and $\beta = 1$. It can be seen that with increase in $g$, evolution of $|E_1|^2$ shows period doubling route to chaos. This is again ascertained in Fig. 5(b), which depicts the largest Lyapunov exponent, $\lambda_{max}$, as a function of the gain/loss parameter.

We observe that with increase in the gain/loss parameter $g$, $\lambda_{max}$ increases rapidly to very high positive values and beyond a certain critical point, it overflows. An overflowing largest Lyapunov exponent implies that two infinitesimally close trajectories are infinitely separated as the system evolves. This occurs when $g$ is chosen in the neighborhood of $g_{th} \approx 0.8814$. Exact calculation of this critical point is not possible. It is known that inclusion of Kerr nonlinearity changes the PT-threshold of a linear system [10]. Thus, we anticipate that $\lambda_{max}$ overflows around the Kerr-nonlinearity induced PT-threshold.

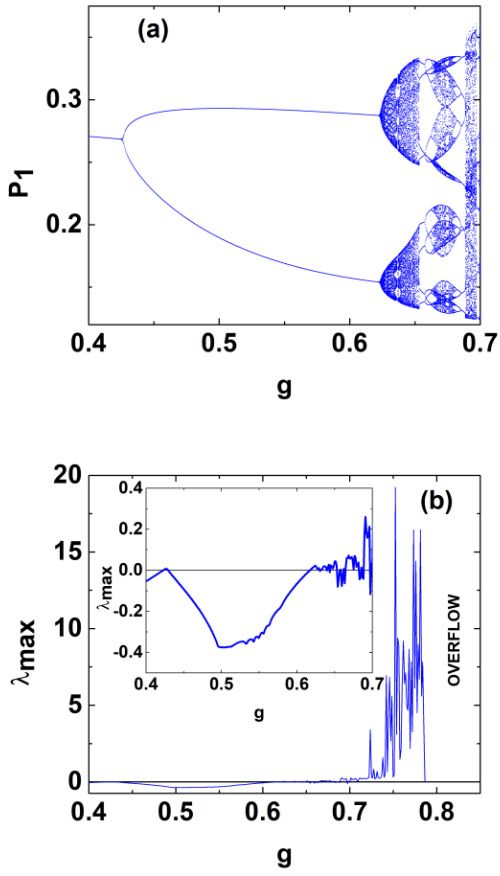

Fig. 5. (a) Bifurcation Diagram and (b) Largest Lyapunov Exponent ($\lambda_{max}$) vs. Gain/Loss Coefficient ($g$)

It is worthwhile to note the significance of high positive value of $\lambda_{max}$. In Fig. 5(b) $\lambda_{max} \approx 19$ corresponds to $g \approx 0.75$. Now, Fig. 6 depicts the evolution of $P_1$ and $P_4$ in the resonator from 1000 to over 3000 iterations for $g = 0.75$. For the chosen parameters, our calculations show that $\lambda_{max} \approx 19.22$. Thus, one could see that optical power increases to extremely high values randomly. This suggests that for specific parametric choices, the evolution of optical power in the resonators will exhibit chaotic spiking.

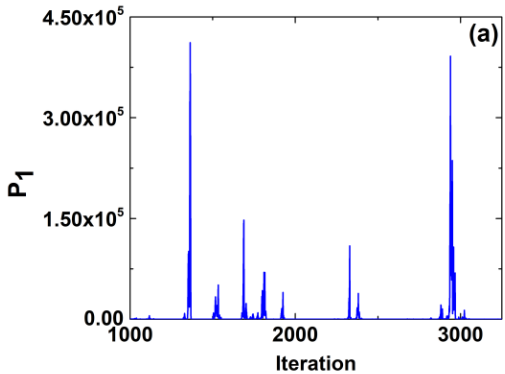

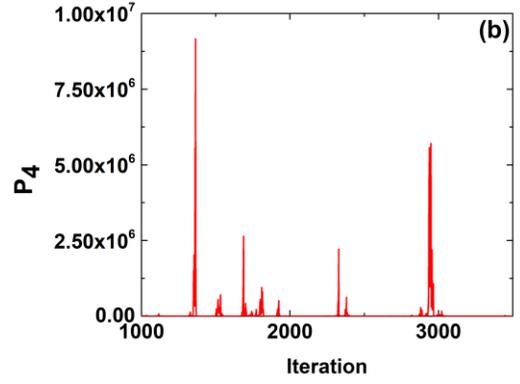

Fig. 6. Evolution of optical power in the two resonators when $g = 0.75$. $E_{in} = 1$, $E'_{in} = 1$.

## 4 Conclusion

We theoretically modeled the *PT* symmetric double ring resonator using the discrete-time iterative equation known as Ikeda Map. In the absence of nonlinearity, we observed optical power saturation below the *PT* threshold and blow-up above the *PT* threshold. We found that optical power saturation occurs in the unbroken *PT* regime because of the existence of stable stationary states on the surface of 4D hypersphere. The hypersphere ceases to exist above the *PT* threshold. When Kerr nonlinearity is taken into account, the emergence of a non-convergent region in the parameter basin is observed. The largest Lyapunov exponent in this region is a real positive quantity, which implies the existence of chaos. Moreover, there exists a divergent region in the parameter basin above the *PT* threshold, where the Lyapunov exponent blows up. The existence of high non-negative largest Lyapunov exponent causes optical power spiking in the resonators.

**Acknowledgements-** J.P.D. would like to thank MHRD, Govt. of India for financial support through a fellowship and A.K.S. would like to acknowledge the financial support from DST-SERB, Government of India (Grant No. SB/FTP/PS-047/2013).

### REFERENCES
[1] C. M. Bender and S. Boettcher: Real Spectra in Non-Hermitian Hamiltonians Having PT Symmetry. Phys. Rev. Lett. **80**, 5243 (1998).
[2] R. El-Ganainy, K.G. Makris, D.N. Christodoulides and Z.H. Musslimani: Theory of coupled optical PT-Symmetric structures. Opt. Lett. **32**, 2632 (2007).
[3] A. Guo *et al.*: Observation of PT-symmetry breaking in complex optical potentials. Phys. Rev. Lett. **103**, 093902 (2009).
[4] C. E. Ruter *et al.*: Observation of parity–time symmetry in optics. Nat. Phys. **6**, 192 (2010).

[5] Xin-You Lü, Hui Jing, Jin-Yong Ma and Ying Wu: PT-Symmetry-Breaking Chaos in Optomechanics Phys. Rev. Lett. **114**, 253601 (2015).
[6] S. K. Gupta and A. K. Sarma: Comm. Nonlin. Sci. Num. Simul. **36**, 141 (2016).
[7] B. Peng *et al.*: Parity–time-symmetric whispering-gallery microcavities. Nat. Phys. **10**, 394 (2014).
[8] M. Duanmu, K. Li, R. L. Horne, P. G. Kevrekidis, N. Whitaker: Linear and nonlinear parity-time-symmetric oligomers: a dynamical systems analysis. Phil. Trans. Roy. Soc. A **371**, 20120171 (2013).
[9] K. Li, P. G. Kevrekidis, D. J. Frantzeeskakis, C. E. Rüter, D. Kip: Revisiting the PT -symmetric trimer: bifurcations, ghost states and associated dynamics. J. Phys. A: Math. Gen. **46**, 375304 (2013).
[10] J. P. Deka and A. K. Sarma: Perturbative dynamics of stationary states in nonlinear parity-time symmetric coupler. Comm. Nonlin. Sci. Num. Simul. **57**, 26 (2017).
[11] V.V. Konotop, J. Yang and D. A. Zezyulin: Nonlinear waves in PT-symmetric systems. Rev. Mod. Phys. **88**, 035002 (2016).
[12] W. Liu et al. An integrated parity-time symmetric wavelength-tunable single-mode microring laser. Nat. Comm. **8**, 15389 (2017).
[13] H. Hodaei, M.-A. Miri, M. Heinrich, D. N. Christodoulides and M. Khajavikhan. Parity-time–symmetric microring lasers. Science **346**, 975(2014).
[14] K. Ikeda, H. Daido, O. Akimoto: Optical Turbulence: Chaotic Behavior of Transmitted Light from a Ring Cavity. Phys. Rev. Lett. **45**, 709 (1980).
[15] K. Ikeda: Multiple-valued stationary state and its instability of the transmitted light by a ring cavity system. Opt. Comm. **30**, 257 (1979).
[16] G. Steinmeyer, D. Jaspert, and F. Mitschke: Observation of a period-doubling sequence in a nonlinear optical fiber ring cavity near zero dispersion. Opt. Commun. **104**, 379 (1994).
[17] M. Haelterman: Period-doubling bifurcations and modulational instability in the nonlinear ring cavity: an analytical study. Opt. Lett. 17, 792 (1992).
[18] G. Steinmeyer, A. Buchholz, M. Hansel, M. Heuer, A. Schwache, and F. Mitschke: Dynamical pulse shaping in a nonlinear resonator. Phys. Rev. A 52, 830 (1995).
[19] R. Vallée: Temporal instabilities in the output of an all-fiber ring cavity. Opt. Commun. 81, 419 (1991).
[20] S. Zaitsev, O. Gottlieb and E. Buks: Nonlinear dynamics of a microelectromechanical mirror in an optical resonance cavity. Nonlinear Dyn. **69**, 1589 (2012).
[21] M. J. Akram and F. Saif: Complex dynamics of nano-mechanical membrane in cavity optomechanics. Nonlinear Dyn. **83**, 963 (2016).
[22] J. P. Deka, S. K. Gupta and A. K. Sarma, Nonlin. Dyn. **87**, 1121, (2017).
[23] S. Lynch, A. L. Steele, J. E. Hoad: Stability analysis of nonlinear optical resonators. Chaos Solitons Fractals **9**, 935–946 (1998)